\documentstyle[12pt]{article}
\def\stacksymbols #1#2#3#4{\def\theguybelow{#2}
\def\verticalposition{\lower#3pt}
\def\spacingwithinsymbol{\baselineskip0pt\lineskip#4pt}
\mathrel{\mathpalette\intermediary#1}}
\def\intermediary#1#2{\verticalposition\vbox{\spacingwithinsymbol
\everycr={}\tabskip0pt
\halign{$\mathsurround0pt#1\hfil##\hfil$\crcr#2\crcr
\theguybelow\crcr}}}

\def\lapproxeq{\stacksymbols{<}{\sim}{2.5}{.2}}
\def\gapproxeq{\stacksymbols{>}{\sim}{3}{.5}}

\begin{document}

\centerline{\bf On the Ionisation of Warm Opaque Interstellar Clouds}
\centerline{\bf and the Intercloud Medium}

\vspace{72pt}

\centerline{D.W. Sciama}

\centerline{SISSA, Strada Costiera 11, 34014, Trieste, Italy}
\centerline{sciama@sissa.it}
\centerline{ICTP, Trieste}
\centerline{Department of Physics, Oxford University}

\newpage

\centerline{\bf Abstract}

In this paper we use a number of observations to construct an integrated
picture of the ionisation in the interiors of quiescent warm opaque interstellar
clouds and in the intercloud medium (ICM) outside dense HII regions and
hot dilute bubbles. Our main conclusion is that within $\sim$ 1kpc of
the sun the ionisation rate of hydrogen per unit volume in both the
interiors of such clouds and in the ICM is independent of the
local density of neutral hydrogen, and varies with position by less than
$\sim$ 20 per cent. These conclusions strongly favour the decaying
neutrino hypothesis for the ionisation of the interstellar medium in
these regions. 

Our analysis is based on a variety of observations, of which the most
remarkable is the discovery by Spitzer and Fitzpatrick (1993) that, in
the four slowly moving clouds along the line of sight to the halo star
HD93521, the column densities of both SII and CII$^*$, which individually
range over a factor $\sim$4, are proportional to the column density of HI
to within $\sim$20 per cent. This proportionality is used to show that
the free electrons exciting the CII to CII$^*$ are located mainly in the
interiors of the clouds, rather than in their skins, despite the large
opacity of the clouds to Lyman continuum radiation. The same conclusion
also follows more unambiguously from the low value of the H$\alpha$ flux
in this direction which was found by Reynolds (1996) in unpublished
observations.

These results are then used, in conjunction with observations of three
pulsar parallaxes and dispersion measures, and with data on HeI, NII and
OI line emissions, to constrain the ionisation of H, He, N and O and the
flux of Lyman continuum photons from O stars in the ICM. It is argued
that the total density of the ICM is kept close to the electron density
in the clouds by a regulation mechanism, and a possible such mechanism
is suggested. This picture of the ionisation of the ICM also favours the
decaying neutrino theory.

Subject headings: Dark matter-elementary particles-Galaxy: halo Galaxy:
solar neighbour-hood-ISM: clouds

\newpage

\centerline{\bf 1. Introduction}

In this paper we present an integrated picture of the processes leading
to the production of low ionisation stages of H, He, C, N, O and S in
warm $(T \sim 10^{4}K)$ and opaque $(N (HI) \sim 10^{19}{\rm cm}^{-2})$
interstellar clouds and in the intercloud medium (ICM) near the sun (d
$\lapproxeq$ 1 kpc) outside dense HII regions and hot dilute bubbles. In
particular we attempt to answer the following questions:

\begin{itemize}
\item[(i)] Are the free electrons in the opaque clouds along the line of
sight to the halo star HD93521, detected by Spitzer and Fitzpatrick
(1993) (SF), located mainly in the cloud skins, where they might be
produced by ionising photons from O stars, or in the cloud interiors
where most of these photons cannot penetrate?

\item[(ii)] Insofar as the hydrogen ionisation process in these regions
is roughly uniform, is the uniform quantity the ionisation rate per unit
volume or per unit hydrogen atom? The first possibility would arise in
an opacity-controlled process involving a uniform production rate of
ionising photons, whereas the second would arise in an essentially
transparent process, such as ionisation by a uniform distribution of
cosmic rays.

\item[(iii)] Is the ICM highly ionised $(n_e/n_{HI} \gg 1)$ or only
partially ionised $(n_e/n_{HI} \lapproxeq 1)$?

\item[(iv)] Why is the free electron density in the ICM, as derived from
pulsar parallaxes and dispersion measures, so similar to that found by SF in
the warm opaque clouds?

\item[(v)] Do the answers to questions (i) to (iv) favour a particular
ionisation mechanism in these regions?
\end{itemize}

We will argue that the correct answers to these questions are the
following:

\begin{itemize}
\item[(i)] The free electrons in the clouds are located mainly in their
interiors, as originally proposed by SF.

\item[(ii)] The ionisation rate in the clouds and in the ICM is indeed
uniform, but per unit volume rather than per unit hydrogen atom.

\item[(iii)] The ICM is highly ionised, as already proposed by Reynolds (1989).

\item[(iv)] The answer to this question is based on a combination of the constancy 
of the ionisation rate per unit volume and the process of cloud
formation, as discussed in section 3.

\item[(v)] These answers strongly favour the hypothesis (Sciama 1990a,
1993a) that the hydrogen ionisation in these regions is mainly due to
photons from decaying dark matter neutrinos.

\newpage

{\bf {2. The Ionisation of Warm Opaque Interstellar Clouds}}
\\
\centerline{2.1. The Slowly Moving Clouds}
\\
Our discussion of the ionisation of these clouds is based on the
remarkable observations of SF, who used the Hubble Space Telescope to
observe the ultraviolet absorption spectrum of the halo star HD93521
$(l=183^0, b=62^0, z \sim 1.5 {\rm kpc})$.  Nine warm clouds with differing velocities
had already been discovered in this direction by Danly et al (1992), who
used HI 21 cm emission measurements with a velocity resolution of 1 km\ 
${\rm s}^{-1}$.  SF observed several species in each of these clouds in 
absorption, and in particular measured the column densities of SII 
and CII$^*$ (the excited J=3/2 state of CII), with a precision varying from
5 to 15 per cent.  They argued that the CII$^*$ is probably excited more by
collisions with free electrons than with neutral hydrogen atoms, and
used their observations to determine the electron density $n_e$ in each
cloud.  The first question which we discuss here is whether these free
electrons are located mainly in the skins of the clouds or in their
interiors.  The resolution of this question should help to determine the
ionisation mechanism involved.  

Our discussion will be simplified if we first restrict ourselves to the
four warm clouds which are moving slowly relative to the local standard
of rest $(\mid v \mid \leq 16.5{\rm km}\ {\rm s}^{-1})$.  Because of the high galactic latitude of the clouds,
their observed velocity needs little correction for galactic rotation,
and we shall assume that the interaction of these slowly moving
clouds with the ICM can be neglected.  As we shall see, these clouds
possess a well-ordered structure in their pattern of column densities.
In section 2.2 we discuss the fast-moving clouds $(v \sim -37\ {\rm to} -65 {\rm km}\ {\rm s}^{-1})$, 
which would be
interacting more strongly with the ICM, and which show a less well-
ordered structure in their pattern of column densities.  In particular,
the ratios N(SII)/N(HI) and N(CII$^*$)/N(HI) do not vary appreciably
amongst the slow clouds, but do vary amongst the fast clouds.  We shall
make considerable use of these facts in our discussion.

We now argue that the free electrons observed in the slow clouds are
located mainly in the interiors of these clouds.  Our arguments are
based on two different considerations, each of which leads to this
conclusion.  In the first, which is an extension of an argument due to
SF, we consider the gas phase abundance of S relative to H in the
clouds, and use the observed constancy of N(SII)/N(HI) and N(CII$^*$)/N(HI)
for the four slowly moving clouds to derive the location of the free
electrons.

In our second argument we make use of some unpublished H $\alpha$ observations
in the direction of the clouds which were obtained by Professor Ron
Reynolds, who kindly made them available to us and gave permission for
them to be referred to here (Reynolds 1996).

We begin by noting the observed values of log N(HI) in the four slow
warm clouds, numbered 6,7,8 and 9.  These values are $19.28 \pm 0.05, 19.36 \pm
0.05, 19.30 \pm 0.06$ and $18.79 \pm 0.23$
respectively.  Thus N(HI) corresponds to the clouds being highly opaque at the
Lyman edge, and ranges over a factor $\sim 3.7$ in those clouds, a result which
will be important later on.  These values are tracked by the values of
log N(SII) in these clouds, which are $14.51 \pm 0.02, 14.66 \pm 0.02, 14.52 \pm 0.02$
and $13.97 \pm 0.06$ respectively.
(There is a small contamination due to the cold cloud numbered 7A, which
does not materially affect our discussion.  In what follows we shall
neglect it.)  Hence - log N(SII)/ N(HI) for the four clouds is $4.77 \pm 0.06,
4.70 \pm 0.06, 4.78 \pm 0.07$ and $4.82 \pm 0.25$ respectively.  Within the mostly 
rather small errors involved
these values are the same for each of the four clouds.  Moreover, as
pointed out by SF, this constant value is essentially equal to the log
of the solar abundance ratio - log N(S)/N(H), which is $4.76 \pm 0.05$
according to Grevesse and Anders (1989).

Following SF, we make the natural interpretation that \\  
(a) the gas phase abundance of S in these clouds is the solar one. \\ 
(b) S is mostly SII in the clouds and H is mostly HI. \\
Domgorgen \& Mathis (1994) have questioned both (a) and (b), and we
discuss (and attempt to refute) their arguments later.  For the moment
we shall adopt (a) and (b), and follow out their implications.

We would in any case expect S to be at least singly ionised in both the
skins and the interiors of the clouds since the ionisation potential of
S is only 10.36 ev, and starlight photons could ionise it at least once
throughout the clouds.  N(SIII) was not measured in the slow clouds, but
in the fast ones SF found that N(SIII)/N(SII) $\sim 0.1$.

For HI and HII the distinction between the skins and interiors of the
clouds is an important one.  We will adopt a simple model in which the
total density n of a cloud is the same at all points of the cloud, in
its skin as well as in its interior, in which H in the skin is
completely ionised down to a depth $l_s$ in the cloud, and in which the
free electron density in the interior of the cloud is uniform.  Then from
(a) and (b) we infer that

$${nl_s \lapproxeq 0.2 n_{HI} l_c,}$$
and
$${n_e \lapproxeq 0.2 n_{HI},} \eqno(1)$$

where $n_e$, $n_{HI}$ and the suffix c refer to the interior of a cloud, the suffix
S refers to its skin, and we have neglected the role of He
ionisation.  It follows that

$${l_s \lapproxeq 0.2 l_c.}$$

We now consider N(CII$^*$) and the implied location of the free electrons.
SF found that in the four slow clouds log N(CII$^*$) is $13.36 \pm 0.03,
13.55 \pm 0.02, 13.42 \pm 0.03$ (not $\pm 0.37$ as given by SF (Fitzpatrick 1996a))
and $12.91 \pm 0.07$ respectively.  Hence in these clouds - log
N(CII$^*$)/N(HI) is $5.92 \pm 0.07, 5.81 \pm 0.06, 5.88 \pm 0.08$ and $5.88
\pm 0.26$ respectively.  Again these values are the same within the
errors.  (In practice SF derived n$_e$ from (CII$^*$)/N(SII), since
N(SII) has a smaller error than N(HI)).

In order to interpret this result we shall assume with SF that the CII$^*$
is produced from CII entirely by electron collisions and that C is
singly ionised in both the skin and interior of each cloud.  Then

$$N(CII^*) = \frac{\zeta}{T^{0.4}} (n^2 l_s + n_e nl_c),\eqno(2)$$

where $\zeta$ contains the gas phase abundance of C in the clouds, the
excitation cross-section and the downward transition probability for
CII$^*$, and where the temperature factor includes the small $(\sim T^{0.1})$
temperature dependence of the excitation cross-section (Blum \& Pradhan
1992).
If the free electrons were located mainly in the skins of the clouds,
and if the gas phase abundance of C is the same in each cloud, the
observed constancy of N(CII$^*$)/N(HI) would imply that

$${\frac{n^2 l_s} {T^{0.4}} \propto\ n_{HI}\ell_c.}\eqno(3)$$

Such a relation would be difficult to interpret.  In fact, since $n^2 l_s$
is proportional to the hydrogen-ionising flux incident on a cloud, one
might expect this quantity, if its value is appreciable, to be inversely
correlated with the column density of the cloud's interior.  It is
unlikely that this argument could be circumvented by a suitably
correlated variation of the carbon abundance from cloud to cloud.

By contrast, if the free electrons were located mainly in the interior
of each cloud, in the sense that, say,

$$n^2 l_s \leq 0.2 n_e nl_c, \eqno(4)$$
then $$nl_s \leq 0.2 n_e l_c,$$

and 

$$l_s \lapproxeq 0.04 l_c,$$

from equation (1).  In this case we would indeed have that

$$N(CII^*) \propto N(HI)$$

if the gas phase abundance of C and $n_e T^{-0.4}$ were the same in each cloud.  
The required constancy of $n_eT^{-0.4}$ can be readily understood if the
ionising photons are produced throughout each cloud, which is opaque to
them.  Then if $\phi$ is the production rate of ionising photons per unit
volume inside a cloud, we would have in ionisation equilibrium and for
the on the spot approximation

$$\phi = \alpha n_e^2, \eqno(5)$$

where $\alpha$ is the recombination co-efficient, excluding recombinations
to the ground state.  Since $\alpha \propto T^{-0.8}$ (Osterbrock 1989) the
constancy of $n_e T^{-0.4}$ from cloud to cloud  would imply that $\phi$ has
the same value in each cloud.  As we shall see later, this
constancy of $\phi$ would be expected if the ionisation of hydrogen inside
the clouds is mainly due to photons from decaying neutrinos.

In this picture the cloud skins play a negligible role.  The opposite is
true in a model proposed by Domgorgen \& Mathis (1994).  They suggested
that the gas phase abundance of S in the clouds is only about half the
solar value.  This assumption would require the column density of free
electrons in each cloud to be comparable to N(HI).  This in turn would
permit the free electrons in each cloud to be located mainly in the
cloud skins where they might be produced by photons from O stars.  In
this picture

$$nl_s \sim n_{HI} l_c$$
and so
$$l_s \sim l_c.$$

Also equation (3) would then imply that $n T^{-0.4}$ is the same in each
cloud.

There are a number of difficulties with this picture.\\
(a) In each cloud $l_s$ must be correlated with $l_c$ in just such a way as
to make N(SII)/N(HI) the same in each cloud to within 20 per cent.\\
(b) This constant value of N(SII)/N(HI) just happens to equal the solar
abundance of S.  However the abundance of S in other regions is found to
be close to the solar value (Fitzpatrick 1996b).\\
(c) The required constancy of $nT^{-0.4}$ is difficult to interpret.  Since
T itself is much the same in each cloud (as judged by the widths of the
HI emission lines) one would require that n is much the same in each
cloud.  This seems unnatural, since N(HI) varies by a factor $\sim 3.7$
amongst the clouds.

We conclude from this discussion that the Domgorgen - Mathis suggestion 
is implausible and that the free electrons detected by SF are
probably located mainly in the interiors of the clouds, as they
originally suggested.  However, it would be desirable to find direct
observational evidence for this conclusion, rather than relying on
indirect arguments resulting from the observed correlation between
N(SII), N(CII$^*$) and N(HI), which might in fact be coincidental.  Such
direct observational evidence does indeed exist, as we now show.  It
involves the H $\alpha$ emission which is associated with the
recombination of the ionised gas in the clouds.  This emission would be
much greater if the ionised gas is in the cloud skins rather than in
their interiors.

To see this we note from equation (2) that if the free elections were
located mainly in the cloud skins the emission measure of the ionised
gas would be proportional to N(CII$^*$).  However, if the free electrons
were located mainly in the cloud interiors, the emission measure would
be reduced by $n_e/n$ which, according to equation (1), is less than $\sim 0.2$.  A
measurement of the H $\alpha$ flux in this direction might then be able to
distinguish between the two models.

To carry out this test in practice we would need to know, at least
approximately, the gas phase abundance of C in the clouds.
Unfortunately the available CII absorption lines were saturated, so this
abundance cannot be measured directly.  Its value depends on both the
total abundance of C and its depletion onto grains.  We know neither of
these quantities for our clouds, although some constraints can be
derived from consideration of the heat balance of the clouds (Sciama
1993b).  Likely values for the interstellar medium near the sun have been
discussed recently by Snow \& Witt (1995, 1996); Cardelli et al (1996);
Sembach \& Savage (1996); Mathis (1996); Kim \& Martin (1996), and Dwek (1997).  
These authors favour a gas
phase abundance of C which is roughly half the solar value.  SF also
adopted this value when they derived $n_e$ from N(CII$^*$).  One must also
worry that the abundance may vary significantly from cloud to cloud.  We shall
neglect this possibiliby here because of the observed constancy of
N(CII$^*$)/N(HI).  We now compare the consequences for the predicted H
$\alpha$ flux of adopting either the solar value or half of it for the
abundance of C.  Later, when we discuss the decaying neutrino theory, we
shall see that the full solar value is preferred by it.

We also need to know the excitation rate for CII$^*$.  This rate is
theoretically uncertain by about 30 per cent (Blum \& Pradhan 1992).
Here we adopt the same parameters as SF.

We now make quantitative estimates of the $H \alpha$ flux expected on each
model.  If the free electrons were located mainly in fully ionised
skins, we would have for the number of ionisations per cm$^2$ per sec
along the line of sight $3\times10^{-8}$ N(CII$^*$) for a C abundance which is
half solar and for T $\sim$ 6000K.  For the four slow clouds the combined N(CII$^*$)
is $9.4\times10^{13} {\rm cm} ^{-2}$, so that for those clouds the total number
of ionisations would be $2.8\times10^6 {\rm cm} ^{-2} s^{-1}$.  If the abundance of
C were solar this number would be $1.4\times10^6 {\rm cm} ^{-2} s^{-1}$.  Similarly
for the five fast clouds together we would have N(CII$^*$) = 2.3$\times10^{13}$
cm $^{-2}$, and so there would be $7\times10^5$ ionisations ${\rm cm}^{-2} s^{-1}$ for a
half solar abundance of C, and $3.5\times10^5 {\rm cm}^{-2} s^{-1}$ for a solar
abundance.

We now compare these predictions with Reynolds' unpublished observations
of the H $\alpha$ flux in this direction for the velocity ranges
corresponding to the two sets of clouds.  For the slow clouds, with the
predicted ionisation rate of $2.8 (1.4)\times10^6 {\rm cm}^{-2} s^{-1}$, the
corresponding H $\alpha$ flux would be 1.4 (0.7) Rayleighs (R).  However,
Reynolds observed for this velocity range only 0.2R in this direction.
Similarly for the fast clouds one would expect an H $\alpha$ flux of 0.35
(0.175) R, whereas for the corresponding velocity range Reynolds
obtained an upper limit of 0.09R.

We regard these substantial discrepancies as demonstrating unambiguously
that the free electrons are not located mainly in fully ionised skins of
the clouds.  As we have seen, however, if they were located mainly in
the interiors of the clouds the predicted H $\alpha$ fluxes would be
reduced by the ratio $n_e/n$ in the interiors.  It is sufficient to assume
that $n_e/n < 1/7 (1/3.5)$ inside the slow clouds to achieve
compatibility with Reynolds' data, with a similar ratio for the fast
clouds.  This constraint is in fact similar to that derived in equation (1) for
the slow clouds from the constancy of N(SII)/N(HI).

Since we will argue in section 4 that the most likely source of the internal 
free electrons is decay photons, we must ensure that the number of such photons 
produced in the ICM is compatible with Reynolds' unusually low value for 
the H $\alpha$ flux in this direction.  More generally we must understand why 
the H $\alpha$ flux and the dispersion measure for distant pulsars yield variations of
$\int n_e^2 ds\ {\rm and}\ \int n_e ds$ by a factor of order 2 in different directions 
(Reynolds 1984, 1991a).  This phenomenon is probably related to the fact that, 
while the local value of $n_e$ in warm opaque regions lies between 0.04 and 
$0.1 {\rm cm}^{-3}$ (depending on the C abundance in the clouds) (see below), the mean 
value of $n_e$ along a line of sight to a pulsar is only $0.033 {\rm cm}^{-3}$
(Nordgren, Cordes \& Terzian 1992).  This difference is presumably due mainly to the 
presence along a typical line of sight of hot $(T \sim 10^6K)$ dilute $(n \sim 10^{-3} 
{\rm cm}^{-3})$ bubbles which are detectable in soft x-rays and are probably produced by 
supernova explosions.  Simple modelling of a stochastic distribution of such bubbles 
easily leads to variations in $\int n_e^2 ds$ and $\int n_e ds$ by a factor of order 2 
along individual long lines of sight.

Having concluded that the free electrons exciting the CII are located mainly 
in the interiors of the clouds, our next task is to derive the value of $n_e$
in the slow clouds.  This derivation was carried out by SF, who showed that, for a half
solar abundance of C, and for $T \sim$ 6000K, $n_e \sim 0.1 {\rm cm}^{-3}$.  For a solar
abundance of C one would, of course, obtain $n_e \sim 0.05 {\rm cm}^{-3}$ (Allowing for the 
various uncertainties involved one could also have $n_e \sim 0.04 {\rm cm}^{-3}.)$.  
This reduction by a factor $\sim$ 2 is important since it would imply, from equation (4), 
a fourfold reduction in the ionisation rate $\phi$ in the clouds (which $\sim$ 
$10^{-15} {\rm cm}^{-3} s^{-1}$ for $n_e \sim 0.05 {\rm cm}^{-3}$ and $T \sim$ 6000K).  We shall
in fact see later that, if $\phi$ is due to decay photons, various observations 
lead to the constraint $n_e \lapproxeq 0.05 {\rm cm}^{-3}$.

\newpage

\centerline{2.2 The Fast-Moving Clouds}

The fast-moving clouds have velocities relative to the LSR of $-64.6, -56.1, -49.5$
and $-37.1 {\rm km}.s^{-1}$.  (The data for the cloud with velocity $-27.4 {\rm km}.s^{-1}$ 
are too inaccurate to be useful).  The scatter in N(CII$^*$)/N(HI) and in N(CII$^*$)/N(SII), and so in 
SF's derived value of $n_e$ for these clouds, is much larger than for the slow-moving
clouds, and its mean is different.  For these reasons we shall treat these clouds
as a separate population, and attribute the differences to effects arising from
the larger impact of the clouds on the ICM, and to possible differences in the gas
phase abundance of C in the clouds.

Following SF we also attempt to explain their observed values of N(SIII)
in the fast clouds, which are about 0.1 of N(SII).  We agree with them
that a flux of photons in the energy range 50-100 ev is the best
explanation for the SIII, but we attribute this flux to impact effects.
(For an alternative view see Benjamin \& Danley 1992).  SF's
discussion shows that the energy flux required $\sim 5\times10^{-7} {\rm erg\ cm}^{-
2}s^{-1}$.  The energy flux available $\sim \frac{1}{2} \rho v^3$, which for an
ICM density $\rho$ of $\sim 10^{-25} {\rm gm\ cm}^{-3}$ and $v \sim 50 {\rm km}\ s^{-1}$
would be $\sim 6\times10^{-6} {\rm erg\ cm}^{-2}s^{-1}$.  Thus one would need about 10\%
efficiency for converting the kinetic energy of the collision into 50-100 ev photons.

Such a flux would also contribute to the ionisation of H in the clouds,
as pointed out by SF.  This ionisation process would have a rate
comparable with that due to the process which we have denoted by $\phi$,
and is presumably at least partly responsible for the increased scatter
in the derived values of $n_e$ for the fast clouds.  Unfortunately it is
not possible to make a clean analysis of the actual values of $n_e$ in
these clouds, since the gas phase abundance of C for them is unknown.
In particular we cannot argue that this abundance is likely to be the
same in each cloud, as we did for the slow clouds which have the same
value of N(CII$^*$)/N(HI).

An alternative explanation for the SIII, and for $n_e$ in all the clouds,
is fossil ionisation (Cox 1995).  There are two difficulties with this
explanation.  The first is that, since SIII recombines to SII about 10
times faster than HII recombines to HI for the same $n_e$, we would expect
that $n_e \sim n_{HI}$ in the clouds, in contradiction to equation (1).
Secondly the observed constancy of $n_e T^{-0.4}$ in the slow clouds would
not be explained.

\end{itemize}

{\bf 3. The Ionisation of the ICM}

In this section we attempt to construct a unified picture of the 
ionisation of the ICM near the sun, but outside dense HII regions and hot dilute bubbles. 
This picture is based on the following observational facts (some of which involve a
certain amount of extrapolation):

\begin{itemize}

\item[a)]In the ICM $n_e \sim 0.04-0.06$cm$^{-3}$ (from pulsar
parallaxes and dispersion measures, see below).
\item[b)]The flux F$_H$ of hydrogen-ionising photons in the ICM does not exceed 
$6\times 10^4 {\rm cm}^{-2}{\rm s}^{-1}$ (from the limits on the emission
measure of cloud skins given by equation (4)).
\item[c)]$n_e/n_{HI} \gapproxeq 10$ for $T \sim 6000$K (from the absence
of OI6300 emission in the ICM (Reynolds 1989)).
\item[d)]$F_{He} < 10^3$cm$^2 s^{-1}$ (from the absence of HeI5875
recombination radiation (Reynolds \& Tufte 1995)).
\item[e)]$F_N > 5 \times 10^3$cm$^{-2}s^{-1}$  for $n_e \sim
0.05$cm$^{-3}$ (from the widespread flux of NII emission in the optical
and the infra-red (Reynolds et al 1973; Reynolds 1991; Bennett et al
1994)).
\end{itemize}

Particularly striking is the result a) which shows that $n_e$ (ICM)
$\sim n_e$ (slow clouds), especially if we adopt the solar value for the
gas phase abundance of C in the clouds, as would be required by
constraints on the decaying neutrino theory (see section 4). The result
a) follows from the data on three pulsars whose distances are known
directly from parallax measurements (Gwinn et al 1986, Bailes et al
1990). The observed dispersion measures of these pulsars combined with
their distances then gives the mean electron density along the line of
sight to each pulsar. For two of the pulsars (PSR 0950+68 and PSR 1451-
68) one must correct for the portions of the lines of sight which pass
through known hot dilute bubbles, in order to derive the mean electron
density along the occupied portions of the lines of sight. One finds in
this way that $n_e \sim 0.056$cm$^{-3}$ for each of these two lines of
sight (Reynolds 1990a; Sciama 1990b). For the third pulsar PSR 0823+26
no correction is needed for an intervening bubble, so the derivation of
$n_e$ along this line of sight is very direct. The distance of this
pulsar is 357$\pm$80pc, and the associated value of $n_e$ is
0.054$\pm$0.012cm$^{-3}$
(Reynolds 1990a). (We note in passing that Reynolds (1990a, 1995) has
emphasised how difficult it is to account for such a degree of ionisation
along this particular line of sight in terms of nearby O or B stars or
white dwarfs). We therefore arrive at a value of $n_e \sim 0.04-
0.06$cm$^{-3}$ in the ICM for these three pulsars.

These results are striking because they tell us that $n_e$ in the ICM is
closely equal to $n_e$ in the slow clouds, especially if the gas phase
abundance of C in the clouds is solar. This equality would be easy to
understand if

\begin{itemize}
\item[(i)]the flux of ionising photons from O stars in the ICM is
significantly less than the flux corresponding to $\phi$, assumed to be
the same in the ICM as in the slow clouds (as it would be in the
decaying neutrino theory).

\item[(ii)]the total gas density in the ICM exceeds $n_e$, since then
$n_e$ would be the same in the ICM as in the slow clouds if the
temperatures in these two regions were similar.
\end{itemize}

However, we wish to accommodate c), which we shall assume to hold
throughout the ICM, and not just in the region of the galactic plane
where Reynold's OI observations were actually conducted. The result c),
in conjunction with b), is telling us that the total gas density in the
ICM is closely equal to $n_e$ in the slow clouds.

We now seek an explanation of this equality, which we prefer not to
regard as a coincidence. A possible explanation might be found along the
following lines. We note that the bulk of the interstellar medium lies
in warm and cold clouds rather than in the ICM (Dickey \& Lockman
1990). Thus cloud formation is an efficient process. The question then
arises, what determines the density $n_i$ of the material which manages
to resist incorporation into clouds and forms the ICM? Since $n_i \sim
n_e \gg n_{HI}$ in the ICM, it seems likely that a key role is played by
the interstellar magnetic field, to which the ionised component of the
ICM is locked. This component is also coupled to the neutral gas by
collisions. It is well known that in many circumstances this results in
the magnetic field lines being frozen into the neutral and ionised
components combined, although there are exceptional cases (Mestel \&
Spitzer 1956). One could imagine that initially, when the gas density
was significantly greater than $n_e$, the mainly neutral interstellar
gas readily formed clouds, despite the resistance due to the stresses
associated with the coupling of the magnetic lines of force to the
ionised component of the gas. However, when the cloud formation process
had proceeded to the point where the total gas density in the ICM had
decreased to $n_e$ in the clouds, that is to $(\phi/
\alpha)^{\frac{1}{2}}$, the neutral hydrogen density in the ICM would
have suddenly dropped substantially, and the relatively small number of
neutral hydrogen atoms remaining in the ICM might no longer be able to
push the ions and the magnetic field lines into clouds. However, the HI
would still be effectively coupled to the HII, that is, ambipolar
diffusion would be a slow process (Mestel \& Spitzer 1956; Spitzer
1978), so that cloud formation would cease, and in this final
configuration one would have $n_i \sim n_e$, as we require.

We now consider the relation between the production rate $\phi$ of
ionising photons in the ICM and the flux $F_{\phi}$ of these photons at
a typical point of the ICM lying between cloud boundaries, when the
density $n$ of the ICM is close to $(\phi/
\alpha)^{\frac{1}{2}}$. This relation is controlled by the opacity due
to the neutral component of density $n_{HI}$, which in turn is
controlled by $F_{\phi}$. Simple modelling shows that the governing
parameter of this problem is the quantity $n \sigma r$ where
$\sigma$ is the photoionisation cross-section at the Lyman edge, and $r$
is the spacing between clouds. For $n_i \sim 0.05$cm$^{-3}$ and $r \sim
20$pc we would have $n_i \sigma r \sim 20$. The models then show
that $n_{HI}$ various more slowly than $r$ (in one model $n_{HI}$
decreased by 25\% when $r$ doubled). We found for a simple model that for
this choice of $n_i$ and $r$, and with $\phi \sim 10^{-15}$cm$^{-3}$s$^{-
1}$, we would have $n_{HI} / n_i \sim 0.1$ and $F_{\phi} \sim 3 \times
10^{4}$cm$^{-2}$s$^{-1}$. In such a model the gas lying between cloud
boundaries would be optically thick at the Lyman edge.

Finally we consider the He and N observational constraints d) and e).
The N constraint is particularly important for us since photons from
decaying neutrinos must have an energy less than 13.8eV and so cannot
ionise N (ionisation potential 14.5eV)(Sciama 1995), and we must appeal
to photons from O stars.

It is difficult to estimate the hydrogen-ionising flux $F^{\prime}$ from
O stars at a typical point in the ICM because of the uncertain porosity
of the distribution of opaque warm and cold clouds. An early attempt was
made by Torres-Peimbert et al (1974) and recent discussions have been
given by Miller \& Cox (1993) and by Dove \& Shull (1994). The He and
N observational constraints act in opposite directions, that is, the
lack of He ionisation in the ICM points to a low O star flux (Reynolds
\& Tufte, 1995), whereas the N constraint requires a certain minimum
value of this flux.

We can satisfy both these requirements, and the other ones which we have
already discussed, if we adopt $F^{\prime} \sim 5 \times 10^3$cm$^{-
2}$s$^{-1}$ and $F_{\phi} \sim 3 \times 10^4$cm$^{-
2}$s$^{-1}$.  Then O stars would produce only $\sim 1/7$ of the total
hydrogen ionising flux $F_{H}$ in the ICM, which is less demanding on
the porosity of the cloud distribution than in the models of Miller \&
Cox and of Dove \& Shull, who required the whole of $F_{H}$ to be due
to $F^{\prime}$. Our relatively low value of $F^{\prime}$ would also
ensure that the electron density in the warm opaque clouds and in the
ICM are similar if their temperatures are similar and if $\phi$ is the
same in both regions.

\newpage

{\bf 4. The Ionisation Source}

We must now try to find a plausible origin for the ionisation source 
$\phi$ inside the four slowly moving warm opaque clouds and in the ICM.
This question for the clouds was discussed by SF, who pointed out a
number of difficulties which a successful mechanism must overcome. These
difficulties are:

\begin{itemize}

\item[(i)]the relatively large value of $n_e$ inside the clouds,
\item[(ii)]the large opacity of the clouds to Lyman continuum photons,
\item[(iii)]the power input required,
\item[(iv)]the quiescence of the slowly moving clouds.
\end{itemize}

In addition to overcoming these difficulties the ionisation source
$\phi$ must possess the following properties:

\begin{itemize}
\item[(a)]It must be the same within each of the four clouds within 20 per
cent.
\item[(b)]It must be the same in the ICM as in the clouds.
\item[(c)]It must be independent of the atomic hydrogen density $n_{HI}$ in
the clouds.
\item[(d)]Its value in the clouds and in the ICM $\sim 6.4 \times
10^{-16} - 1.4 \times 10^{-15}$cm$^{-3}$s$^{-1}$ in order to produce an
electron density $n_e \sim 0.04-0.06$cm$^{-3}$ for T$\sim 6000$K.
\end{itemize}

SF were concerned to understand how the ionisation could occur inside
the clouds despite their large opacity at the Lyman edge. They
considered shock waves, cosmic rays and soft X-rays as possible sources.
Cox (1995) added the suggestion, which we have already referred to, that
the clouds might not be in ionisation equilibrium, but might be steadily
recombining after one or more local explosive events had fully ionised
them.

Insofar as $n_e$ in the clouds is comparable to its value in the ICM,
one problem with all SF's explanations is an energetic one. If the
ultimate energy source for the shock waves, cosmic rays or soft X-rays
were supernova explosions then, as Reynolds (1990b) emphasised, nearly
100 per cent of the available supernova kinetic energy would have to be
channelled into maintaining the ionisation of the interstellar medium. Moreover these
processes would be expected to lead to an ionisation rate proportional
to $n_{HI}$ rather than independent of it, so leading to a varying
$n_{e}$ in the clouds in the absence of an unnatural coincidence.
Fossil ionisation would also be expected to lead to an unacceptably
large ionisation ratio and to a varying $n_e$.

By contrast, the decaying neutrino theory immediately solves all these
difficulties and actually requires properties a) - c)to hold, while d) is
numerically reasonable, as we show below. In this theory $\phi =
n_{\nu}/ \tau$, where $n_{\nu}$ is the local neutrino number density and
$\tau$ is the decay lifetime.  So long as the clouds and the relevant 
portions of the ICM are closer
together than the scale-length of the neutrino distribution one would
expect $\phi$ to be approximately constant. In fact HD93521 lies at a
distance of about 1.7kpc and about 1.5kpc above the plane of the Galaxy
(Spitzer \& Fitzpatrick 1992). We do not know where the slow clouds are
situated along the line of sight, but it is reasonable to place them at
heights below $\sim 1$kpc (Benjamin \& Danly 1997). In addition the
pulsars with known parallaxes all lie less than 400pc from the sun.
These distances are less than the expected scale-length of the neutrino
distribution. Moreover the flattened halo models of Binney, May \&
Ostriker (1987) show that the pinch effect of the disk on the halo
reduces its density by only 10 per cent at a distance of 500pc above the
plane. Hence properties a) and b) are predictions of the decaying
neutrino theory.

The same is true of property c) since, as we have seen, in an opaque
region in ionisation equilibrium $\phi = \alpha n^2_e$, a relation which
is independent of $n_{HI}$ (so long as the opacity condition is
satisfied).

Finally we consider property d) which tells us the required value of
$\phi$. To see whether this value is reasonable for the decaying
neutrino theory we note that the scale height of the ICM $\sim$ 700pc
(Reynolds 1991b; Nordgren et al 1992). The neutrinos in the layer within
this height would produce $\sim 2 \times 10^6$ ionisations cm$^{-2}$s$^{-
1}$ which agrees with the value derived from Reynold's (1984) $H \alpha$
data. The neutrinos lying outside this gas layer would also contribute
ionising photons, but this contribution is reduced by the familiar
factor 4 arising from the integration over solid angle associated with a
transparent region bordered by an opaque slab.

We also note that a lower limit to the value of $\tau$ can be derived from
observational data which constrain the isotropic extragalactic
background at 2000$A$. The value of this background is still
controversial (Bowyer 1991; Wright 1992; Henry 1991; Henry \& Murthy 1993; Witt \&
Petersohn 1994; Witt, Friedmann \& Sasseen 1997), but despite this uncertainty one can deduce that $\tau \geq
10^{23}$s (Sciama 1991; Overduin, Wesson \& Bowyer 1993; Dodelson \& Jubas 1994); otherwise the
redshifted background due to the cosmological distribution of neutrinos
would exceed the upper limit permitted by analysis of the observed
background at 2000A.

With this observational constraint on $\tau$ it follows that the required value of $\phi$
implies that $n_{\nu} \geq 6.4 \times 10^7 - 1.4 \times 10^8$cm$^{-3}$.
Now in this theory the mass of the decaying neutrino $\sim$ 28eV
(Sciama 1993a), and so the mass density of neutrinos near the sun $\gapproxeq 0.04-0.1$M$_{\odot}$pc$^{-3}$.

This value must not exceed estimates of the amount
 of dark matter near the sun derived from the z velocities of nearby stars
(the problem of the Oort limit). These estimates are also still
controversial (Kuijken \& Gilmore 1991; Kuijken 1991; Bahcall, Flynn \&
Gould 1992). We follow Binney et al (1987), who constructed models of
flattened dark halos of the Galaxy and compared the model densities near
the sun with constraints derived from observed stellar z velocities. One
of their acceptable models, with a halo flattening $\sim 0.3$, leads to
a halo dark matter density near the sun $\sim 0.03$M$_{\odot}$pc$^{-3}$.
Our required value of $\gapproxeq$ 0.04-0.1M$_{\odot}$pc$^{-3}$ is in essential
agreement with this model at the lower end of its range.  We conclude
that the decaying neutrino theory can give rise to the required value of
$\phi$ if $\tau$ $\sim$ 10$^{23}$ s.

\newpage

{\bf 5. Conclusions}

In this paper the main points we have made are the following:

\begin{itemize}
\item[(1)]Four warm opaque slowly moving interstellar clouds near the
sun are observed to contain a significant density $n_e$ of free
electrons in their interiors, with $n_e \sim 0.04-0.1$cm$^{-3}$,
depending on the gas phase abundance of C in the clouds.
\item[(2)]The quantity $n_e T^{-0.4}$ is the same in each cloud to a
precision of 10 per cent.
\item[(3)]The value of $n_e$ in the ICM near the sun, outside dense HII
regions and dilute hot bubbles, $\sim 0.04-0.06$cm$^{-3}$.
\item[(4)]In the ICM $n_{HI}/n_e \lapproxeq 0.1$.
\item[(5)]The only plausible explanation of all these results taken
together is that dark matter neutrinos in the Galaxy are producing
hydrogen-ionising photons at a constant rate $\phi$ near the sun $\sim
10^{-15}$cm$^{-3}$s$^{-1}$, and that the total density $n_i$ of the ICM is
regulated so that $n_i \sim (\phi / \alpha)^{\frac{1}{2}}$. A possible
regulation mechanism is suggested.
\item[(6)]The high abundance of NII and the low abundance of HeII in the
ICM would be explained if at a typical point near the sun the flux of
O star photons in the Lyman continuum is about $\frac{1}{6}$ of the flux
of decay photons. The actual fraction depends on the porosity of the
distribution of opaque warm and cold clouds near the sun, and cannot be
estimated a priori with any precision.
\end{itemize}

We are grateful to R.J. Reynolds for informing us of his unpublished
H$\alpha$ observations, for giving us permission to refer to them, and
for helpful and encouraging discussions. We are also grateful to the
Italian Ministry for Universities and Scientific and Technological Research for
financial support.

\newpage

\def\ref{\parskip=0pt\par\noindent\hangindent\parindent
    \parskip=2ex plus 0.5ex minus 0.1ex}

\centerline{\bf References}

\vskip 1cm

{\parskip 0pt 
\hangindent=.5cm\hangafter=1
\noindent Bahcall, J.N., Flynn, C. \& Gould, A., 1992, {\it ApJ}, {\bf 389},
234 

\hangindent=.5cm\hangafter=1
\noindent Bailes, M., Manchester, R.N., Kesteven, M.J., Norris, R.P. \&
Reynolds, J.E., 1990, {\it Nature}, {\bf 343}, 240

\hangindent=.5cm\hangafter=1
\noindent Benjamin, R.A. \& Danly, L., 1997, to be published

\hangindent=.5cm\hangafter=1
\noindent Bennet, C.L., et al., 1994 {\it ApJ}, {\bf 434}, 587

\hangindent=.5cm\hangafter=1
\noindent Binney, J.J., May, A. \& Ostriker, J.P., 1987, {\it MNRAS}, {\bf
226}, 149

\hangindent=.5cm\hangafter=1
\noindent Blum, R.D. \& Pradhan, A.K., 1992, {\it
ApJS}, {\bf 80}, 425

\hangindent=.5cm\hangafter=1
\noindent Bowyer, S., 1991, {\it ARA\&A}, {\bf 29}, 59

\hangindent=.5cm\hangafter=1
\noindent Cardelli, J.A., Meyer, D.M., Jura, M. \& Savage, B.D., 1996, {\it
ApJ}, {\bf 467}, 334

\hangindent=.5cm\hangafter=1
\noindent Cox, D.P., 1995, in The Physics of the Interstellar Medium and the
Intergalactic Medium ASP Conf. Series Vol. 80, ed. A. Ferrara, C.F. McKee,
C. Heiles, \& P.R. Shapiro, (ASP. San Francisco) 317 

\hangindent=.5cm\hangafter=1
\noindent Danly, L., Lockman, F.J., Meade, M.R., \& Savage, B.D., 1992, {\it
ApJS}, {\bf 81}, 125

\hangindent=.5cm\hangafter=1
\noindent Dickey, J.M. \& Lockman, F.J., 1990, {\it A.R.A. \& A.}, {\bf 28}, 215

\hangindent=.5cm\hangafter=1
\noindent Dodelson, S. \& Jubas, J.M., 1994, {\it MNRAS}, {\bf 266}, 886 

\hangindent=.5cm\hangafter=1
\noindent Domgorgen, H., \& Mathis, J.S., 1994, {\it ApJ}, 428, 647

\hangindent=.5cm\hangafter=1
\noindent Dove, J.B. \& Shull, J.M., 1994, {\it ApJ}, {\bf 430}, 222

\hangindent=.5cm\hangafter=1
\noindent Dwek, E., 1997, {\it astro-ph9701109}

\hangindent=.5cm\hangafter=1
\noindent Fitzpatrick, E., 1996a, private communication

\hangindent=.5cm\hangafter=1
\noindent Fitzpatrick, E., 1996b, {\it ApJ}, {\bf 473}, L55

\hangindent=.5cm\hangafter=1
\noindent Gwinn, C.R., Taylor, J.H., Weisberg, J.M. \& Rawlings, L.A., 1986,
{\it ApJ}, {\bf 91}, 338

\hangindent=.5cm\hangafter=1
\noindent Grevesse, N., \& Anders, E., 1989, in Cosmic Abundances of Matter,
ed. C.J. Waddington, (New York: AIP), 1

\hangindent=.5cm\hangafter=1
\noindent Henry, R.C., 1991, {\it ARA\&A}, {\bf 29}, 89

\hangindent=.5cm\hangafter=1
\noindent Henry, R.C. \&Murthy, J., 1993 {\it ApJ}, {\bf 418}, L17

\hangindent=.5cm\hangafter=1
\noindent Kim, S.-H. \& Martin, P.G., 1996, {\it ApJ}, {\bf 462}, 296

\hangindent=.5cm\hangafter=1
\noindent Kuijken, K., 1991 {\it ApJ}, {\bf 376}, 467

\hangindent=.5cm\hangafter=1
\noindent Kuijken, K. \& Gilmore, G., 1991, {\it ApJ}, {\bf 367}, L9

\hangindent=.5cm\hangafter=1
\noindent Mathis, J.S., 1996, {\it ApJ}, {\bf 472}, 643

\hangindent=.5cm\hangafter=1
\noindent Mestel, L. \& Spitzer, L., 1956, {\it MNRAS}, {\bf 116}, 503

\hangindent=.5cm\hangafter=1
\noindent Miller, W.W. \& Cox, D.P., 1993, {\it ApJ}, {\bf 417}, 579

\hangindent=.5cm\hangafter=1
\noindent Nordgren, T., Cordes, J. \& Terzian, Y., 1992, {\it AJ}, {\bf 104},
1465

\hangindent=.5cm\hangafter=1
\noindent Osterbrock, D.E., 1989, Astrophysics of Gaseous Nebulae and Active
Galactic Nuclei (Mill Valley, CA: Univ. Science Books)

\hangindent=.5cm\hangafter=1
\noindent Overduin, J.M., Wesson, P.S. \& Bowyer, S., 1993, {\it ApJ}, {\it
404}, 460

\hangindent=.5cm\hangafter=1
\noindent Reynolds, R.J., 1984, {\it ApJ}, {\bf 282}, 181

\hangindent=.5cm\hangafter=1
\noindent Reynolds, R.J., 1989, {\it ApJ}, {\bf 345}, 811

\hangindent=.5cm\hangafter=1
\noindent Reynolds, R.J., 1990a, {\it ApJ}, {\bf 348}, 153

\hangindent=.5cm\hangafter=1
\noindent Reynolds, R.J., 1990b, {\it ApJ}, {\bf 349}, L17

\hangindent=.5cm\hangafter=1
\noindent Reynolds, R.J., 1991a, {\it ApJ}, {\bf 372}, L17

\hangindent=.5cm\hangafter=1
\noindent Reynolds, R.J., 1991b, in IAU Symposium No. 144, The Interstellar
Disk-Halo Connection in Galaxies, ed. H. Bloemen (Dordrecht: Kluwer),
67

\hangindent=.5cm\hangafter=1
\noindent Reynolds, R.J., 1992, {\it ApJ}, {\bf 392}, L35

\hangindent=.5cm\hangafter=1
\noindent Reynolds, R.J., 1995, in The Physics of the Interstellar Medium and
the Intergalactic Medium, ASP Conf. Series Vol. 80, ed. A. Ferrara, 
C.F. McKee, C. Heiles \& P.R. Shapiro, (ASP. San Francisco) 388

\hangindent=.5cm\hangafter=1
\noindent Reynolds, R.J., 1996, private communication

\hangindent=.5cm\hangafter=1
\noindent Reynolds, R.J., Roesler, F.L. \& Scherb, F., 1973 {\it ApJ}, {\bf
179}, 651

\hangindent=.5cm\hangafter=1
\noindent Reynolds, R.J. \& Tufte, S.L., 1995, {\it ApJ}, {\bf 439}, L17

\hangindent=.5cm\hangafter=1
\noindent Sciama, D.W., 1990a,{\it ApJ}, {\bf 364}, 549

\hangindent=.5cm\hangafter=1
\noindent Sciama, D.W., 1990b, {\it Nature}, {\bf 346}, 40

\hangindent=.5cm\hangafter=1
\noindent Sciama, D.W., 1991 in The Early Observable Universe from Diffuse
Backgrounds, ed. B. Rocca-Volmerange, J.M. Deharveng \& J. Trans Thanh
Van, (Edition Frontieres), p.127

\hangindent=.5cm\hangafter=1
\noindent Sciama, D.W., 1993a, Modern Cosmology and the Dark Matter Problem
(Cambridge University Press)

\hangindent=.5cm\hangafter=1
\noindent Sciama, D.W., 1993b, {\it ApJ}, {\bf 409}, L25

\hangindent=.5cm\hangafter=1
\noindent Sciama, D.W., 1995, {\it ApJ}, {\bf 448}, 667

\hangindent=.5cm\hangafter=1
\noindent Sembach, K.R. \& Savage, B.D., 1996, {\it ApJ}, {\bf 457}, 211

\hangindent=.5cm\hangafter=1
\noindent Snow, T.P. \& Witt, A.N., 1995, {\it Science}, {\bf 270}, 1455

\hangindent=.5cm\hangafter=1
\noindent Snow, T.P. \& Witt, A.N., 1996, {\it ApJ}, {\bf 468}, L65

\hangindent=.5cm\hangafter=1
\noindent Spitzer, L., 1978, Physical Processes in the Interstellar Medium
(Wiley, New York), p.294

\hangindent=.5cm\hangafter=1
\noindent Spitzer, L. \& Fitzpatrick, E.L., 1992, {\it ApJ}, {\bf 391}, L41

\hangindent=.5cm\hangafter=1
\noindent Spitzer, L., \& Fitzpatrick, E.L., 1993, {\it ApJ}, {\bf 409}, 299

\hangindent=.5cm\hangafter=1
\noindent Torres-Peimbert, S., Lazcano-Araujo, A. \& Peimbert, M., 1974, {\it
ApJ}, {\bf 191}, 401

\hangindent=.5cm\hangafter=1
\noindent Witt, A.N.,, Friedman, B.C. \& Sasseen, T.P., 1997, {\it
astro-ph9701017}

\hangindent=.5cm\hangafter=1
\noindent Witt, A.N. \& Petersohn, J.K., 1994, in ASP Conf. Ser. 58, 1st
Symp. Infrared Cirrus and Diffuse Interstellar Clouds, ed. R.M. Cutri \&
W.B. Latter, (San Francisco: ASP), 91

\hangindent=.5cm\hangafter=1
\noindent Wright, E.L., 1992, {\it ApJ}, {\bf 391}, 34 

}

\end{document}